\documentclass{elsart}
\usepackage{epsfig}

\newcommand{\beq}{\begin{equation}}
\newcommand{\eeq}{\end{equation}}
\newcommand{\beqa}{\begin{eqnarray*}}
\newcommand{\eeqa}{\end{eqnarray*}}
\newcommand{\ea}{{\it et al.}}

\newcommand{\la}{\mathrel{\hbox{\rlap{\hbox{\lower7pt\hbox{$\sim$}}}\hbox{$<$}}}}
\newcommand{\ga}{\mathrel{\hbox{\rlap{\hbox{\lower7pt\hbox{$\sim$}}}\hbox{$>$}}}}

\begin{document}
\begin{frontmatter}
\title{The Age of the Universe} 
\author[ad1]{Brian Chaboyer}
\address[ad1]{Hubble Fellow, Steward Observatory, The University of
Arizona, \cty Tucson, AZ, \cny USA 85721~~~ e-mail: chaboyer@as.arizona.edu}

\begin{abstract}
A minimum age of the universe can be estimated directly by determining
the age of the  oldest objects in the our Galaxy.  These objects are
the metal-poor stars in the halo of the Milky Way.  Recent work on
nucleochronology finds that the oldest stars are $15.2\pm 3.7$ Gyr
old.  White dwarf cooling curves have found a minimum age for the
oldest stars of 8 Gyr.  Currently, the best estimate for the age of
the oldest stars is based upon the absolute magnitude of the main
sequence turn-off in globular clusters.  The oldest globular clusters
are $11.5\pm 1.3$ Gyr, implying a minimum age of the universe of 
$t_{\rm universe} \ge 9.5$ Gyr (95\% confidence level). 
\end{abstract}

%\begin{keyword}
%Globular clusters \sep  
%Cosmology \sep 
%Stellar structure \sep  
%White dwarfs
%\PACS 97.10.C \sep 97.20.Rp \sep 98.20.Gm \sep 98.80 \sep 98.80.Bp 
%\end{keyword}
\end{frontmatter}

\section{Introduction}
A direct estimate for the minimum age of the universe may be obtained
by determining the age of the oldest objects in the Milky Way.  This
direct estimate for the age of the universe can be used to constrain
cosmological models, as the expansion age of the universe is a simple
function of the Hubble constant, average density of the
universe and the cosmological constant.  The
oldest objects in the Milky Way are the metal-poor stars located in
the spherical halo.  There are currently three independent methods used to
determine the ages of these stars: (1) nucleochronology, (2) white
dwarf cooling curves and (3) main sequence turn-off ages based upon
stellar evolution models.  In this review I will summarize recent
results from these three  methods, with particular emphasize on main
sequence turn-off ages as they currently provide the most reliable
estimate for the age of the universe.

%\cite{JKE,MCTK,KTTK}. 

\section{Nucleochronology}
Conceptually, the simplest way to determine the age of a star is to
use the same method which have been used to date the Earth --
radioactive dating.  The age of a star is derived using the abundance
of a long lived radioactive nuclei with a known half-life (see, for
example the review \cite{Cowan}). The difficulty in applying
this method in practice is the determination of the original abundance
of the radioactive element.  The best application of this method to
date has been on the very metal-poor star CS 22892 \cite{Sneden}.
This star has a measured thorium abundance (half-life of 14.05 Gyr),
and just as importantly, the abundance of the elements from $56 \le Z
\le 76$ are very well matched by a scaled solar system
$r$-process\footnote{The $r$-process is the creation of elements
heavier than Fe through the {\it rapid} capture of neutrons by a seed
nuclei.} abundance distribution.  Thus, it is logical to assume that
the original abundance of thorium in this star is given by the scaled
solar system $r$-process thorium abundance. A detailed study of the
$r$-process abundances in CS 22892 lead to an age of $15.2\pm
3.7\,$Gyr for this extremely metal-poor star \cite{Sneden}.  This in
turn, implies a $2\,\sigma$ lower limit to the age of the universe of
$t_{\rm universe} \ge 7.8\,\,{\rm Gyr}$ from nucleochronology.  This
is not a particularly stringent constraint at present.  However, the
uncertainty in the derived age is due entirely to the uncertainty in
the determination of the thorium abundance in CS 22892.  The
determination of the abundance of thorium in a number of stars with
similar abundance patterns to CS 22892 will naturally lead to a
reduction in the error.  If 8 more stars are observed, then the error
in the derived age will be reduced to $\pm 1.2\,$Gyr, making
nucleochronology the preferred method of obtaining the absolute ages
of the oldest stars in our galaxy.

\section{White Dwarf Cooling Curves}
White dwarfs are the terminal stage of evolution for 
stars less massive than $\sim 8\,{\rm M}_\odot$.  As 
white dwarfs age, they become cooler and fainter.  Thus, the 
luminosity of the faintest white dwarfs can be used to estimate
their age.  This age is based upon theoretical white dwarf cooling
curves \cite{Wood,Segretain,Salaris}.  There are a number of uncertainties
associated with theoretical white dwarf models, which have been
studied in some detail.  However, the effect of these {\it
theoretical} uncertainties are generally not included in deriving the
uncertainty associated with white dwarf cooling ages.  

The biggest difficulty in using white dwarfs to estimate the age of
the universe is that white dwarfs are very faint and so are very
difficult to observe.  Most studies of white dwarf ages have
concentrated on the solar neighborhood, in an effort to determine the
age of the local disk of the Milky Way.  Even these nearby samples can
be affected by completeness concerns.  The  age determination for 
these disk white dwarfs is complicated by the fact that the results
are sensitive to the star formation rate as a function of time \cite{Wood}.  A
recent study has increased the sample size of local white dwarfs and
concluded that the local disk of the Milky Way has an age of  $t_{\rm disk} =
9.5^{+1.1}_{-0.8}\,$Gyr, where the quoted errors are due to the
observational uncertainties in counting faint white dwarfs
\cite{Oswalt}.  This implies a $2\,\sigma$ lower limit to the age of
the local disk of
$
t_{\rm disk} \ge 7.9\,{\rm Gyr}.
$

Recently, with the Hubble Space Telescope it has become possible to
observe white dwarfs in nearby globular clusters\footnote{Globular
clusters are compact stellar systems containing $\sim 10^5$ stars.
These stars contain few heavy elements (typically 1/10 to 1/100 the
ratio found in the Sun) and are spherically distributed about the
Galactic center.  Together, these facts suggest that globular clusters 
were among the first objects formed in the Galaxy.  There is evidence
for an age range among the globular clusters, so the tightest limits on
the minimum age of the universe are found when only the oldest
globular clusters are considered.  These are typically the globular
clusters with the lowest heavy element abundances (1/100 the solar ratio).}.  
These observations are not deep enough to
observe the faintest white dwarfs and  can only put a lower limit
to the age of the white dwarfs.  Observations of the globular cluster
M4 found a large number of white dwarfs, with no decrease in the
number of white dwarfs at the faintest observed magnitudes
\cite{Richer}.  Based upon the luminosity of the faintest observed
white dwarfs, a lower limit to the age of M4 was determined to be
$t_{\rm glob} \ga 8\,$Gyr \cite{Richer}.  When the advanced camera
becomes operational on HST (scheduled to occur in the year 2000), it
will be possible to obtain considerably deeper photometry of M4,
leading to an improved constraint on the age of M4 from white dwarf
cooling curves.

\section{Main Sequence Turn-off Ages}
\begin{figure}
\centerline{\epsfig{file=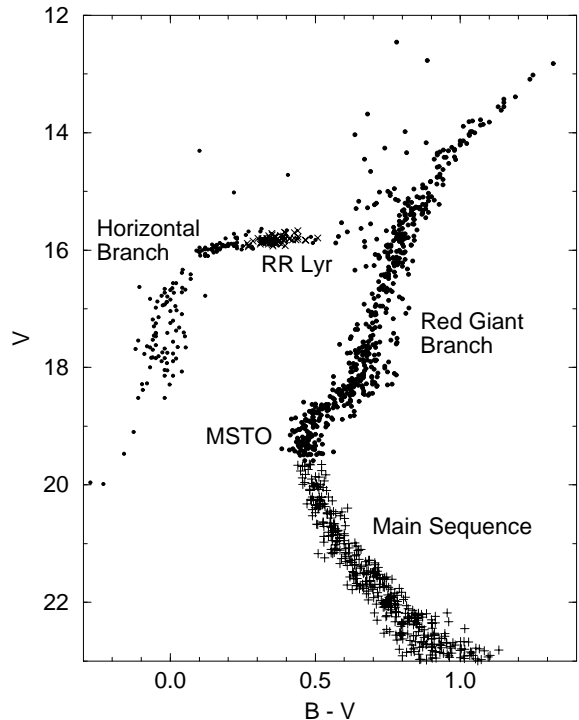,height=9.0cm}}
\caption{A color-magnitude diagram of a typical globular cluster, M15
[10].  
The vertical axis plots the magnitude (luminosity) of the stars in the
V wavelength region, with brighter stars having smaller magnitudes.
The horizontal axis plots the color (surface temperature) of the
stars, with cooler stars towards the right.  All of the stars in a
globular cluster have the same age and chemical composition.  Their
location in the color-magnitude diagram is determined by their mass.
Higher mass stars have shorter lifetimes and evolve more quickly than
low mass stars.  The various evolutionary sequence have been labeled.
Most stars are on the main sequence (MS), fusing hydrogen into helium
in their cores (for clarity, only about 10\% of the stars on the MS
have been plotted).  Slighter higher mass stars have exhausted their
supply of hydrogen in the core, and are in the main sequence turn-off
region (MSTO). After the MSTO, the stars quickly expand, become
brighter and are referred to as red giant branch stars (RGB).  These
stars are burning hydrogen in a shell about a helium core.  Still higher
mass stars have developed a helium core which is so hot and dense that
helium fusion is ignited.  This evolutionary phase is referred to as
the horizontal branch (HB).  Some stars on the horizontal branch are
unstable to radial pulsations.  These radially pulsating variable
stars are called RR Lyrae stars, and are important distance indicators.
}
\label{cmd}
\end{figure}

Theoretical models for the evolution of stars provide an independent
method to determine stellar ages.  These computer models are based on
stellar structure theory, which is outlined in numerous textbooks
\cite{schw,hanson}.  One of the triumphs of stellar evolution theory
is a detailed understanding of the preferred location of stars in a
temperature-luminosity plot (Figure \ref{cmd}).
 
A stellar model is constructed by solving the four basic equations of
stellar structure: (1) conservation of mass; (2) conservation of
energy; (3) hydrostatic equilibrium and (4) energy transport via
radiation, convection and/or conduction.  These four, coupled
differential equations represent a two point boundary value problem.
Two of the boundary conditions are specified at the center of the star
(mass and luminosity are zero), and two at the surface.  In order to
solve these equations, supplementary information is required.  The
surface boundary conditions are based on stellar atmosphere
calculations.  The equation of state, opacities and nuclear reaction
rates must be known.  The mass and initial composition of the star
need to be specified.  Finally, as convection can be important in a
star, one must have a theory of convection which determines when a
region of a star is unstable to convective motions, and if so, the
efficiency of the resulting heat transport.  Once all of the above
information has been determined a stellar model may be constructed.
The evolution of a star may be followed by computing a static stellar
structure model, updating the composition profile to reflect the
changes due to nuclear reactions and/or mixing due to convection, and
then re-computing the stellar structure model.

There are a number of uncertainties associated with stellar evolution
models, and hence, age estimates based on the models.  Probably the
least understood aspect of stellar modeling is the treatment of
convection.  Numerical simulations hold promise for the
future \cite{abbett,kim}, but at present one must view properties of
stellar models which depend on the treatment of convection to be
uncertain, and subject to possibility large systematic errors. Main
sequence, and red giant branch globular cluster stars have surface
convection zones. Hence, the surface properties of the stellar models
(such as its effective temperature, or color) are rather uncertain.
Horizontal branch stars have convective cores, so the predicted
luminosities and lifetimes of these stars are subject to possible
systematic errors.

\begin{figure}
\centerline{\epsfig{file=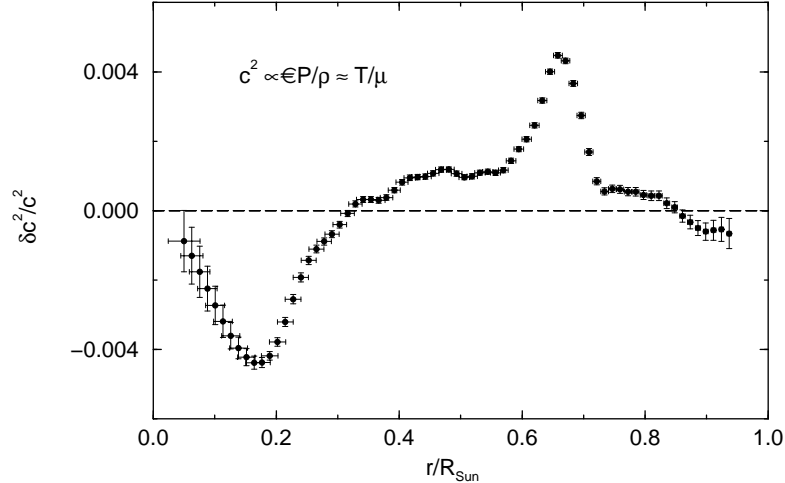,height=10.0cm,angle=270}}
\caption{The difference between the squared sound speed between a
theoretical solar model and the actual Sun ($[c^2_{\rm model} -
c^2_{\rm Sun}]/c^2_{\rm model}$), as a function of radius of the
model.  The sound speed of the Sun is obtained from helioseismology
--- observations of the solar $p$-modes, whose frequencies depend on
the sound speed.  Note that the maximum difference between the model
and the Sun is less than 0.5\%, a level of accuracy rarely seen in
astronomy.  The best solar models constructed in the early 1990's lead
to squared sound speed differences of order 3\%.}
\label{inversion}
\end{figure}

Given the known uncertainties in the models, the luminosity (absolute
magnitude) of the main-sequence turn-off has the smallest theoretical
errors, and is the preferred method for obtaining the absolute ages of
globular clusters (e.g.\ \cite{renzini,dm96}).  The theoretical
calibration of age as a function of the luminosity of the
main-sequence turn-off has changed somewhat over the last several
years.  It has long been realized that diffusion (the settling of
helium relative to hydrogen) could shorten the predicted main sequence
lifetimes of stars \cite{noerd80}.  However, it was not clear if
diffusion actually occurred in stars, so this process had been ignored
in most calculations.  Recent helioseismic studies of the Sun have 
shown that diffusion occurs in the Sun
\cite{dalsgaard,guenther}.  The Sun is a typical main sequence star,
whose structure (convective envelope, radiative interior) is quite
similar to main sequence globular cluster stars.  Thus, as diffusion
occurs in the Sun, it appears likely that diffusion also occurs in
main sequence globular cluster stars.  Modern calculations find that
the inclusion of diffusion lowers the age of globular clusters by 7\%
\cite{chab2}.  The recent use of an improved equation of state has led
to a further 7\% reduction in the derived globular cluster ages
\cite{chabeos}. The equation of state now includes the effect of
Coulomb interactions \cite{rogers}.  Helioseismic studies of the Sun
find that there are no significant errors associated with the equation
of state currently used in stellar evolution calculations
\cite{basu}. Together, the use of an improved equation of state and
the inclusion of diffusion in the theoretical
models have lead to a $\sim 2$ Gyr (14\%) reduction in the estimated
ages of for the oldest globular clusters.  The excellent agreement
between theoretical solar models and the Sun (see Figure
\ref{inversion}) suggest that future improvements in stellar models
will likely lead to small (less than $\sim 5\%$) changes in the
derived ages of globular cluster stars.  

A detailed Monte Carlo study found that the uncertainties in the
theoretical models led to an 1-$\sigma$ error of 7\% in the derived
globular cluster ages \cite{chaboyer}.  This study considered errors
associated with 15 different parameters used in the construction of
theoretical stellar models and isochrones.  The parameter which lead
to the largest uncertainty in the derived age of the globular clusters
was the abundance of the $\alpha$-capture elements (oxygen is the most
important $\alpha$-capture element) in globular cluster stars.  It is
difficult to determine the abundance of oxygen observationally
\cite{nissen,bala}, with estimates of the oxygen abundance varying by
up to a factor of 3. When extreme values for the oxygen abundance are
used in the theoretical calculations, the derived globular cluster
ages change by 8\%.

The use of the luminosity of the main sequence turn-off as an age
indicator requires that the distance to the globular cluster be known.
Determining distances is one of the most difficult tasks in astronomy,
and is always fraught with uncertainty.  The release of the Hipparcos
data set of parallaxes to nearby stars \cite{hipp} has suggested that
a revision in the conventional globular cluster distance scale is
necessary.  Hipparcos did not directly determine the distance to any
globular clusters, but did provide the distance to a number of nearby
metal-poor main sequence stars.  Assuming that globular cluster stars
have identical properties to these nearby stars, the nearby stars can
serve as calibrators of the intrinsic luminosity of metal-poor main
sequence stars and the distance to a globular cluster determined.
This technique is referred to as main sequence fitting.  There have
been a number of papers which have used the Hipparcos data set to
determine the distance to globular clusters using main sequence
fitting \cite{chaboyer,reid,gratton,pont}.  Three of these papers
conclude that globular clusters are further away than previously
believed, leading to a reduction in the derived ages.  The remaining
paper \cite{pont} concluded that the Hipparcos data did not lead to a
revision in the globular cluster distance scale.  However, this work
incorrectly included binary stars in the main sequence fitting
\cite{chaboyer}. When the known binaries are removed from the fit (a
case which is also considered in \cite{pont}), then all four papers
are in agreement --- the Hipparcos data yields larger distances (and
hence, younger ages) for globular clusters.  My analysis
\cite{chaboyer} considered four distance determination techniques in
addition to using the Hipparcos data, and concluded that the five
independent distance estimates to globular clusters {\em all\/} led to
younger globular cluster ages.

A number of authors have recently examined the question of the
absolute age of the oldest globular clusters.  All of these works used
the luminosity of the main sequence turn-off as the age indicator.
The results are summarized in Table \ref{agetable}.  Despite the fact
that these investigators used a variety of theoretical stellar models
(with differing input physics) and different methods to determine the
distance to the globular clusters, the derived ages are remarkably
similar, around 12 Gyr.  These ages are $\sim  3$ Gyr younger than
previous determinations, due to improved input physics used in the
models, and a longer distance scale to globular clusters. 
My work \cite{chaboyer} considered a variety of distance
indicators and included a very detailed Monte Carlo study of
the possible errors associated with the theoretical stellar models.
For this reason, my preferred age for the oldest globular clusters is
$11.5\pm 1.3$ Gyr, implying a minimum age of the universe of $t_{\rm
universe} \ge 9.5\,\, {\rm Gyr}$ at the 95\% confidence level.
\begin{table}
\caption{Estimates for the age of the oldest globular clusters}
\label{agetable}
\begin{center}
\begin{tabular}{llcl}
\hline\hline
\multicolumn{1}{c}{Age (Gyr)}&
\multicolumn{1}{c}{Distance determination}& 
\multicolumn{1}{c}{Reference}\\
\hline
~~\\[-14pt]
$11.5\pm 1.3$  &  5 independent techniques & \cite{chaboyer}\\
$12\pm 1$ & main sequence fitting (Hipparcos) & \cite{reid}\\
$11.8\pm 1.2$  & main sequence fitting (Hipparcos)  & \cite{gratton}\\
$14.0\pm 1.2$ & main sequence fitting (Hipparcos) including binaries& \cite{pont}\\
$12\pm 1$ & theoretical HB \& main sequence fitting & \cite{dantona}\\
$12.2 \pm 1.8$ &  theoretical HB & \cite{sal2}\\
~~\\[-14pt]
\hline
\end{tabular}
\end{center}
\end{table}

\section{Summary}
A direct estimate for the minimum age of the universe can be obtained
by determining the age of the oldest objects in the galaxy.  These
objects are the metal-poor stars located in the halo of the Milky Way.
There are currently three independent techniques which have been
used to determine the ages of the metal-poor stars in the Milky Way:
nucleochronology, white dwarf cooling theory, and main sequence
turn-off ages.  The best application of nucleochronology to date has
been on the very metal-poor star CS 22892 which has an age of $15.2\pm
3.7\,$Gyr \cite{Sneden}, implying a $2\,\sigma$ lower limit to the age
of the universe of $t_{\rm universe} \ge 7.8\,\,{\rm Gyr}$.  White
dwarf cooling theory is difficult to apply in practice, as one needs
to observe very faint objects. Currently, it is impossible to observe
the faintest white dwarfs in a globular cluster, so white dwarf
cooling theory can only provide a lower limit to the age of a globular cluster.
Based upon the luminosity of the faintest observed white dwarfs, a
lower limit to the age of M4 was determined to be $t_{\rm glob} \ga
8\,$Gyr \cite{Richer}.  Absolute globular cluster ages based upon the
main sequence turn-off have recently been revised due to a realization
that globular clusters are farther away than previously thought.  The
age of the oldest globular clusters is $11.5\pm 1.3$ Gyr
\cite{chaboyer}, implying a minimum age of the universe of $t_{\rm
universe} \ge 9.5$ Gyr (95\% confidence level).  At the present time,
main sequence turn-off ages have the smallest errors of the available
age determination techniques and provide the best estimate for the age
of the universe.

To obtain the actual age of the universe, one must add to the above
age the time it took for the metal-poor stars to form.
Unfortunately, a good theory for the onset of star formation within
the galaxy does not exist.  Estimates for the epoch of initial star
formation range from redshifts of $z \sim 5$ to 20.  This corresponds
to ages ranging from 0.1 to 2 Gyr, implying that the actual age of
the universe lies in the range $9.6 \le t_{\rm universe} \le
15.4\,$Gyr. Tightening the bounds of this estimate will require a
better understanding of the epoch of galaxy formation, along with
improved stellar models and distance estimates to globular clusters.

\section*{Acknowledgment}   
The author was supported for this work by NASA through Hubble
Fellowship grant number HF--01080.01--96A awarded by the Space
Telescope Science Institute, which is operated by the Association of
Universities for Research in Astronomy, Inc., for NASA under contract
NAS 5--26555.

\end{document}